\documentclass[12pt]{article}
\usepackage{amsfonts}
\usepackage{amsmath}
\usepackage{amssymb}
\usepackage{graphicx}
\usepackage{color}
\usepackage[all, knot]{xy}
\usepackage{tikz}
\usepackage{array}

\usepackage{ulem}

\usepackage[utf8]{inputenc}
\usepackage{epstopdf}
\usepackage[footnotesize]{caption}
\usepackage{amsthm}
\usepackage{enumitem}
\usepackage{mathrsfs}

\usepackage[margin=3cm]{geometry}

\def \be {\begin{equation}}
\def \ee {\end{equation}}
\def \bea {\begin{eqnarray}}
\def \eea {\end{eqnarray}}
\def \nn {\nonumber}

\def \rr {\raise.35ex\hbox{\small $\prime$}\kern-.17em{\mbox{\large $\imath$}}}

\def \dels {\partial\kern-.6em /\kern.1em}
\def \As {{A\kern-.5em / \kern.5em}}
\def \Ds {D\kern-.7em / \kern.5em}

\def \ks {k\kern-.5em /}
\def \ls {l\kern-.5em /}

\newcommand{\ci}[1]{}

\newcommand{\ba}{\begin{eqnarray}}
\newcommand{\ea}{\end{eqnarray}}
\newcommand{\bal}{\begin{align}}
\newcommand{\eal}{\end{align}}
\newcommand{\bay}[1]{\left(\begin{array}{#1}}
\newcommand{\eay}{\end{array}\right)}

\newcommand{\hide}[1]{}

\newlist{axioms}{enumerate}{2}
\setlist[axioms,1]{label=\textbf{A\arabic{axiomsi}.}, ref=A\arabic{axiomsi}}
\setlist[axioms,2]{label=\textbf{A\arabic{axiomsi}\rlap{\myEnumCounter{axiomsii}}.},%
                   ref=A\arabic{axiomsi}\myEnumCounter{axiomsii},%
                   align=parleft,%
                   leftmargin=0em,%
                   itemsep=1.4ex,%
                   before={\stepcounter{axiomsi}}}

  \usetikzlibrary{decorations.markings}

\begin{document}
\begin{titlepage}

\begin{center}

\textbf{\LARGE
Quantifying Quantum Entanglement in\\ 
Two-Qubit Mixed State from\\ 
Connected Correlator
\vskip.3cm
}
\vskip .5in
{\large
Xingyu Guo$^{a, b}$ \footnote{e-mail address: guoxy@m.scnu.edu.cn}
and Chen-Te Ma$^{a, b, c, d, e, f}$ \footnote{e-mail address: yefgst@gmail.com}
\\
\vskip 1mm
}
{\sl
$^a$
Guangdong Provincial Key Laboratory of Nuclear Science,\\
 Institute of Quantum Matter,
South China Normal University, Guangzhou 510006, Guangdong, China.
\\
$^b$
Guangdong-Hong Kong Joint Laboratory of Quantum Matter,\\
 Southern Nuclear Science Computing Center, 
South China Normal University, Guangzhou 510006, Guangdong, China.
\\
$^c$ 
Department of Physics and Astronomy, Iowa State University, Ames, Iowa 50011, US.
\\
$^d$
Asia Pacific Center for Theoretical Physics, \\
Pohang University of Science and Technology, 
Pohang 37673, Gyeongsangbuk-do, South Korea. 
\\
$^e$
School of Physics and Telecommunication Engineering,\\ 
South China Normal University, Guangzhou 510006, Guangdong, China.
\\
$^f$
The Laboratory for Quantum Gravity and Strings,\\
 Department of Mathematics and Applied Mathematics,\\
University of Cape Town, Private Bag, Rondebosch 7700, South Africa. 
}
\\
\vskip 1mm
\vspace{40pt}
\end{center}
\newpage
\begin{abstract}
Our study employs a connected correlation matrix to quantify Quantum Entanglement. 
The matrix encompasses all necessary measures for assessing the degree of entanglement between particles. 
We begin with a three-qubit state and involve obtaining a mixed state by performing partial tracing over one qubit. 
Our goal is to exclude the non-connected sector by focusing on the connected correlation. 
This suggests that the connected correlation is deemed crucial for capturing relevant entanglement degrees. 
The study classifies mixed states and observes that separable states exhibit the lowest correlation within each class. 
We demonstrate that the entanglement measure monotonically increases concerning the correlation measure. 
This implies that connected correlation serves as an effective measure of Quantum Entanglement. 
Finally, our proposal suggests that interpreting Quantum Entanglement from a local perspective is possible. 
The observable is described as a vector with locality but violates freedom of choice. 
\end{abstract}
\end{titlepage}

\section{Introduction}
\label{sec:1}
\noindent 
In quantum information theory, the separability problem refers to determining whether a given mixed state can be expressed as a statistical mixture of product states \cite{Werner:1989zz}. 
For two-qubit systems, the Peres-Horodecki criterion, also known as the Positive Partial Transpose (PPT) criterion \cite{Peres:1996dw,Horodecki:1996nc}, provides a necessary and sufficient condition for separability. 
It involves taking the partial transpose of the density matrix $\rho^{T_2}$, where The $T_2$ is a partial transpose operation on region 2, and checking the positivity of the resulting eigenvalues. 
If all eigenvalues are non-negative, the state is separable; otherwise, it is entangled. 
\\

\noindent
Negativity is a measure derived from the PPT criterion. 
It is the absolute value of the sum of the negative eigenvalues of the partial transposed density matrix. 
In other words, it is the sum of the absolute values of the negative eigenvalues minus their actual values, divided by 2: 
\bea
N\equiv\bigg|\sum_{j, \lambda_j<0}\lambda_j\bigg|=\sum_j\frac{|\lambda_j|-\lambda_j}{2}, 
\eea
where $\lambda_j$ is the eigenvalue of the partial transposed density matrix. 
Negativity is useful in distinguishing separable states from entangled states in two-qubit mixed states that entanglement entropy cannot. 
It has been used in cosmology to infer cosmic acceleration in entangled states between two epochs \cite{Capozziello:2013wea}. 
The quantification of three-qubit pure state Quantum Entanglement is {\it not} enough if losing some measures \cite{Guo:2021ghs,Guo:2021kvq}. 
The result suggests that negativity might not capture all aspects of entanglement in such cases. 
We expect that negativity cannot have a monotonic result in the correlation. 
Therefore, the negativity should {\it not} show the amount of Quantum Entanglement even in two-qubit mixed states. 
It helps solve the separability problem in a two-qubit mixed state but may {\it not} be a complete measure of entanglement in more complex systems. 
The distinction between the separability problem (determining if a state can be expressed as a mixture of product states) and the quantification problem (determining the degree of entanglement) is crucial. 
Negativity primarily addresses the former.
\\

\noindent 
Bell's inequality provides a test for local hidden variable theories \cite{Bell:1964kc}. 
Violation of Bell's inequality indicates the failure of a {\it local realistic} description, realism (existence before measuring), and {\it freedom to the choice} between measurement settings (which implies that the probability distribution of the hidden variable is independent of measure settings) \cite{Bell:1964kc,Clauser:1969ny,Hensen:2015ccp}. 
In Bell's inequality, the correlation between two separated particles is measured by a classical correlation function denoted as $c$ and defined by the formula
\bea
c(\vec{a}, \vec{b})\equiv\int dP(\lambda)\ E(\vec{a}, \vec{b}, \lambda),       
\eea
where $P$ is a probability distribution, and $E(\vec{a}, \vec{b}, \lambda)$ is an observable. 
This function depends on a local hidden variable $\lambda$ and represents the average correlation. 
The locality condition assumes that the measurement outcomes on the particles, represented by unit vectors $\vec{a}$ and $\vec{b}$, are independent and can be factorized
\bea
\label{sep}
E(\vec{a}, \vec{b}, \lambda)=E(\vec{a}, \lambda)E(\vec{b}, \lambda),  
\eea   
If Bell's inequality $|c(\vec{a}, \vec{b})+c(\vec{a}, \vec{b^{\prime}})+c(\vec{a^{\prime}}, \vec{b})-c(\vec{a^{\prime}}, \vec{b^{\prime}})|\le 2$ is violated, it implies the presence of entanglement in the two-qubit state. 
To quantify entanglement, the quantum correlator $c_{\mathrm{Q}}\equiv\mathrm{Tr}(\rho\vec{a}\cdot\vec{\sigma}\otimes\vec{b}\cdot\vec{\sigma})$ is considered, which recovers the experimental result \cite{Cirelson:1980ry}. 
The $\vec{\sigma}\equiv(\sigma_x, \sigma_y, \sigma_z)$ is the vector of the Pauli matrix. 
The $\rho$ is a density matrix. 
The {\it violation} of locality condition results in $c_{\mathrm{Q}}$ not following Bell's inequality, as the quantum correlator cannot be factorized into two scalar functions. 
The maximum violation degrees (of Bell's theorem) are monotonically increasing with the concurrence \cite{Bennett:1996gf,Wootters:1997id,Verstraete:2001} 
\bea
C(\psi)\equiv\sqrt{2(1-\mathrm{Tr}\rho_{\mathrm{R}}^2)}.
\eea 
The reduced density matrix is denoted by $\rho_{\mathrm{R}}$. 
The maximum violation ($\gamma$) is given by the two {\it largest} eigenvalues ($u_1, u_2$) of $R^TR$ \cite{Verstraete:2001}, where 
\bea
R_{jk}\equiv\mathrm{Tr}(\rho\sigma_j\otimes\sigma_k), \ \gamma=2\sqrt{u_1+u_2}.
\label{mv}
\eea 
The concurrence and the maximum violation degree are introduced as measures of entanglement. 
Three-qubit states \cite{Acin:2000jx} are classified into two inequivalent entangled classes \cite{Dur:2000zz}. 
The two-body entanglement measures are not enough. 
Genuine tripartite-measures, like the three-tangle, are introduced for three-body entanglement \cite{Coffman:1999jd}. 
However, the two largest eigenvalues of the generalized R-matrix (or three-point correlation matrix) work and quantify Quantum Entanglement \cite{Guo:2021ghs,Guo:2021kvq}. 
We want to continue this approach to the two-qubit mixed state. 
The central question that we address in this letter is: {\it How to diagnose Quantum Entanglement in a two-qubit mixed state using the connected correlation matrix?}  
\\

\noindent
In this letter, we investigate the study of mixed states obtained by partial tracing over one qubit from a general three-qubit pure state. 
Our primary objective is to gain a deeper understanding of the properties of mixed states through the modification of Bell's operator, with a specific emphasis on utilizing the connected part of the correlation matrix. 
Our efforts have led to an intriguing discovery: the maximum violation of Bell's theorem, corresponding to the modified Bell's operator, demonstrates a consistent and monotonic increase concerning the entanglement measures for each classification under consideration. 
Intriguingly, our observations reveal that separable states, those devoid of entanglement, consistently exhibit the lowest correlations within each classification. 
This finding underscores the fundamental connection between entanglement and the violation of Bell's theorem. 
Furthermore, our exploration has led us to identify a classical model that contravenes the locality condition of Bell's theorem through the use of a vector observable. 
It is noteworthy that, despite this violation, the model offers a local interpretation for the vector observable. 
This highlights the inherent challenge in establishing a definitive link between entanglement and non-local correlations, as the classical model manages to incorporate a local interpretation. 
It is imperative to emphasize that these violations we have uncovered do not adhere to the principle of freedom of choice. 
This aspect raises important questions regarding the interplay between entanglement, non-local correlations, and the constraints imposed by the freedom of choice in the context of Bell's theorem.  

\section{Two-Qubit Entanglement Measures}
\label{sec:2}
\noindent 
A general three-qubit quantum state (up to single-qubit unitary transformations) is \cite{Acin:2000jx} 
\bea
|\psi\rangle
&=&
\lambda_0|000\rangle
+\lambda_1e^{i\phi}|100\rangle
+\lambda_2|101\rangle
\nn\\
&&
+\lambda_3|110\rangle
+\lambda_4|111\rangle, 
\eea
where the $\lambda_j$ are non-negative, and the range of $\phi$ is $0\le\phi\le\pi$. 
The normalization of a density matrix $\mathrm{Tr}\rho=1$ provides $\lambda_0^2+\lambda_1^2+\lambda_2^2+\lambda_3^2+\lambda_4^2=1$. 
Therefore, five measures are needed to describe Quantum Entanglement. 
We perform a partial trace over one qubit to obtain the two-qubit mixed states, $\rho_{12}$, $\rho_{13}$, and $\rho_{23}$. 
They have the same entangled degrees with the thee qubit quantum state. 
The following are the three-qubit entanglement measures, and we rewrite them to the measures of $\rho_{12}$, $E_1, C_1, C_2, C_{12}, E_5$, 
\bea
\label{measure12}
E_1&\equiv&\tau_{1|2}=2\lambda_0\lambda_3; 
\nn\\ 
E_2&\equiv&\tau_{1|3}=\sqrt{C_{12}^2+E_1^2-C_2^2}=2\lambda_0\lambda_2; 
\nn\\
E_3&\equiv&\tau_{2|3}=\sqrt{C_{12}^2+E_1^2-C_1^2}=2|\lambda_1\lambda_4e^{i\phi}-\lambda_2\lambda_3|; 
\nn\\
E_4&\equiv&\tau_{1|23}-\tau_{1|2}-\tau_{1|3}
\nn\\
&=&\sqrt{C_1^2+C_2^2-C_{12}^2-2E_1^2} 
=2\lambda_0\lambda_4; 
\nn\\
E_5&\equiv&\mathrm{Tr}\big((\rho_1\otimes\rho_2)\rho_{12}\big)
-\frac{1}{3}\mathrm{Tr}(\rho_1^3)
-\frac{1}{3}\mathrm{Tr}(\rho_2^3)
\nn\\
&&
+\frac{1}{4}(E_1^2+E_2^2+E_3^2+E_4^2)
\nn\\
&=&
\mathrm{Tr}\big((\rho_2\otimes\rho_3)\rho_{23}\big)
-\frac{1}{3}\mathrm{Tr}(\rho_2^3)
-\frac{1}{3}\mathrm{Tr}(\rho_3^3)
\nn\\
&&
+\frac{1}{4}(E_1^2+E_2^2+E_3^2+E_4^2)
\nn\\
&=&
\mathrm{Tr}\big((\rho_3\otimes\rho_1)\rho_{31}\big)
-\frac{1}{3}\mathrm{Tr}(\rho_3^3)
-\frac{1}{3}\mathrm{Tr}(\rho_1^3)
\nn\\
&&
+\frac{1}{4}(E_1^2+E_2^2+E_3^2+E_4^2)
\nn\\
&=&\lambda_0^2(\lambda_2^2\lambda_3^2
-\lambda_1^2\lambda_4^2
+|\lambda_1\lambda_4e^{i\phi}-\lambda_2\lambda_3|^2).  
\eea
The $E_1, E_2, E_3$ are the entanglement of formation for the different subregions, given by \cite{Bennett:1996gf}
\bea
\min_{p_j, \psi_j}\sum_jp_jC(\psi_j). 
\eea 
The $E_4$ is the three-tangle \cite{Coffman:1999jd}. 
The $E_5$ shows the correlation of the reduced density matrices. 
For the two-qubit mixed state, the concurrences are all independent due to that $C_1\neq C_2$ and $C_{12}\neq 0$, where $C_1$ is the concurrence associated with the reduced density matrix $\rho_1$. 
As shown in Eq. \eqref{measure12}, we can interpret the three-qubit entanglement measures using the two-qubit measures. 
For the $\rho_{13}$ ($\rho_{23}$), we choose the measures $E_2, C_1, C_3, C_{13}, E_5$ ($E_3, C_2, C_3, C_{23}, E_5$) and show the similar results as in Eq. \eqref{measure12}. 
Therefore, we can use $E_1, E_2, E_3, E_4, E_5$ to measure Quantum Entanglement in a two-qubit mixed state.  

\section{R-Matrix}
\label{sec:3}
\noindent 
We can observe the R-matrix (3$\times$ 3 matrix) naturally from the Bell's operator \cite{Bell:1964kc,Verstraete:2001}:
\bea
{\cal B}_{\mathrm{Q}}
&=&
c_{\mathrm{Q}}(\vec{a}, \vec{b})
+c_{\mathrm{Q}}(\vec{a}, \vec{b^{\prime}})
+c_{\mathrm{Q}}(\vec{a^{\prime}}, \vec{b})
-c_{\mathrm{Q}}(\vec{a^{\prime}}, \vec{b^{\prime}})
\nn\\
&=&\sum_{j ,k=1}^3
\big(\vec{a}_jR_{jk}(\vec{b}_k+\vec{b^{\prime}}_k)
+\vec{a^{\prime}}_jR_{jk}(\vec{b}_k-\vec{b^{\prime}}_k)\big).
\eea 
The maximum value of $\langle{\cal B}_{\mathrm{Q}}\rangle$ for all choices of $\vec{a}$, $\vec{b}$, $\vec{a^{\prime}}$, and $\vec{b^{\prime}}$ is conventionally called maximum violation \cite{Verstraete:2001}. 
The calculation is equivalent to computing the two largest eigenvalues of the R-matrix \eqref{mv} \cite{Verstraete:2001}. 
Since the R-matrix is a three by three matrix, the eigenvalues ($\lambda$) of $R^TR$ follow the cubic equation 
\bea
\lambda^3+\alpha_1\lambda^2+\alpha_2\lambda+\alpha_3=0.
\eea 
The discriminant is $\Delta\equiv\gamma_1^2+\gamma_2^3$, 
where 
\bea
\gamma_1\equiv-\frac{\alpha_1^3}{27}
-\frac{\alpha_3}{2}
+\frac{\alpha_1\alpha_2}{6}; \ 
\gamma_2\equiv\frac{\alpha_2}{3}-\frac{\alpha_1^2}{9}. 
\eea
Because the eigenvalues are real-valued, the discriminant satisfies $\Delta\le 0$. 
We can rewrite all measures in terms of $E_1, E_2, E_3, E_5$ as in the following: 
\bea
\alpha_1&=&E_2^2+E_3^2-2E_1^2-1; 
\nn\\
\alpha_2&=&(E_2^2-E_1^2)(E_3^2-E_1^2)-8\bigg(E_5-\frac{E_1^2}{4}\bigg); 
\nn\\
\alpha_3&=&-16\bigg(E_5-\frac{E_1^2}{4}\bigg)^2.
\eea
The analytical solution of the maximum violation ($\gamma$) is 
\bea
\gamma=2\sqrt{-\frac{2\alpha_1}{3}+2\sqrt{-\gamma_2}\cos\bigg(\theta-\frac{\pi}{3}\bigg)}, 
\eea
where
\bea
0\le\theta\equiv\frac{1}{3}\arccos\bigg(\frac{\gamma_1}{(-\gamma_2)^{\frac{3}{2}}}\bigg)\le\frac{\pi}{3}.  
\eea
Therefore, the value of $\gamma$ no longer relies on the three-tangle ($E_4$). 
In other words, the quantum correlator cannot diagnose Quantum Entanglement. 

\section{Connected R-Matrix}
\label{sec:4}
\noindent 
Now we replace the quantum correlator with the connected correlator given by
\bea
c_{\mathrm{C}}(\vec{a}, \vec{b})&\equiv&\mathrm{Tr}(\rho\vec{a}\cdot\vec{\sigma}\otimes\vec{b}\cdot\vec{\sigma})
\nn\\
&&
-\mathrm{Tr}(\rho\vec{a}\cdot\vec{\sigma}\otimes I_2)\times\mathrm{Tr}(\rho I_2\otimes\vec{b}\cdot\vec{\sigma}),  
\eea
where $I_2$ is a two-by-two identity matrix. 
The connected R-matrix is 
\bea
R_{c, jk}=\mathrm{Tr}(\rho\sigma_j\otimes\sigma_k)
-\mathrm{Tr}(\rho\sigma_j\otimes I_2)\times\mathrm{Tr}(\rho I_2\otimes\sigma_k). 
\nn\\
\eea
When considering a separable pure state, the connected R-matrix vanishes. 
The general two-qubit pure state shows the maximal connected violation (given by the two largest eigenvalues of $R_c^TR_c$), $\gamma_c=2\sqrt{2}C(\psi)$. 
The monotonically increasing result remains. 
When Quantum Entanglement disappears, the connected correlation vanishes. 
When using $c_{\mathrm{Q}}$, the non-connected correlation provides 2 for $C(\psi)=0$. 
Hence we should remove some unnecessary correlations. 
However, the correlation function is non-factorizable for the separable mixed state. 
Therefore, the connected correlation contributes to Classical and Quantum Entanglement. 
Even so, we will show that this approach quantifies Quantum Entanglement.   
\\

\noindent 
When considering $\rho_{12}$, the $\alpha_1, \alpha_2, \alpha_3$ for $c_{\mathrm{C}}$ are: 
\bea
\alpha_1&=&-(E_1^2+E_2^2+E_4^2)(E_1^2+E_3^2+E_4^2)+2(4E_5-E_1^2); 
\nn\\
\alpha_2&=&E_1^2(E_1^2+E_4^2)(2E_1^2+E_2^2+E_3^2+2E_4^2)
\nn\\
&&
+(4E_5-E_1^2)^2; 
\nn\\
\alpha_3&=&-E_1^4(E_1^2+E_4^2)^2. 
\eea
Now we can find that $\gamma_c$ depends on all necessary entanglement measures, $E_1, E_2, \cdots, E_5$. 
The $\gamma_c$ is monotonic to $-\alpha_1$ with fixed parameters, $\gamma_2$ and $\theta$. 
Exchanging $E_1^2$ and $E_2^2$ ($E_3^2$) shows the result of $\rho_{13}$ ($\rho_{23}$). 
All cases show a similarly monotonic behavior. 
\\

\noindent 
We only turn on one entanglement measure first. 
The proper correlation measure is monotonic for the entanglement quantity. 
We list the general result of $\rho_{12}$ for each single measure, $E_1, E_2\cdots, E_4$: 
\begin{itemize}
\item{
$\lambda_2=\lambda_4=0$, $\gamma_c=2\sqrt{2}E_1$;  
}
\item{
$\lambda_3=\lambda_4=0$, $\gamma_c=0$;
}
\item{
$\lambda_0=0$, $\gamma_c=0$;
}
\item{
$\lambda_1=\lambda_2=\lambda_3=0$, $\gamma_c=2E_4^2$. 
}
\end{itemize}
The $E_5$ must couple to other measures and cannot appear alone. 
The vanishing entanglement of formation implies separability (in a two-qubit state) \cite{Bennett:1996gf}. 
In other words, the density matrix is separable for the cases of a single measure, $E_2, E_3, E_4$. 
For the case of $E_4$, the $\gamma_c$ is not zero and is monotonic for $E_4^2$. 
Therefore, it reflects the fact that $\gamma_c$ measures the classical and quantum correlation simultaneously. 
Our result respects the expectation. 
Other reduced density matrices have a similar result. 
\\  

\begin{figure}[!htb]
\begin{center}
\includegraphics[width=0.32\textwidth]{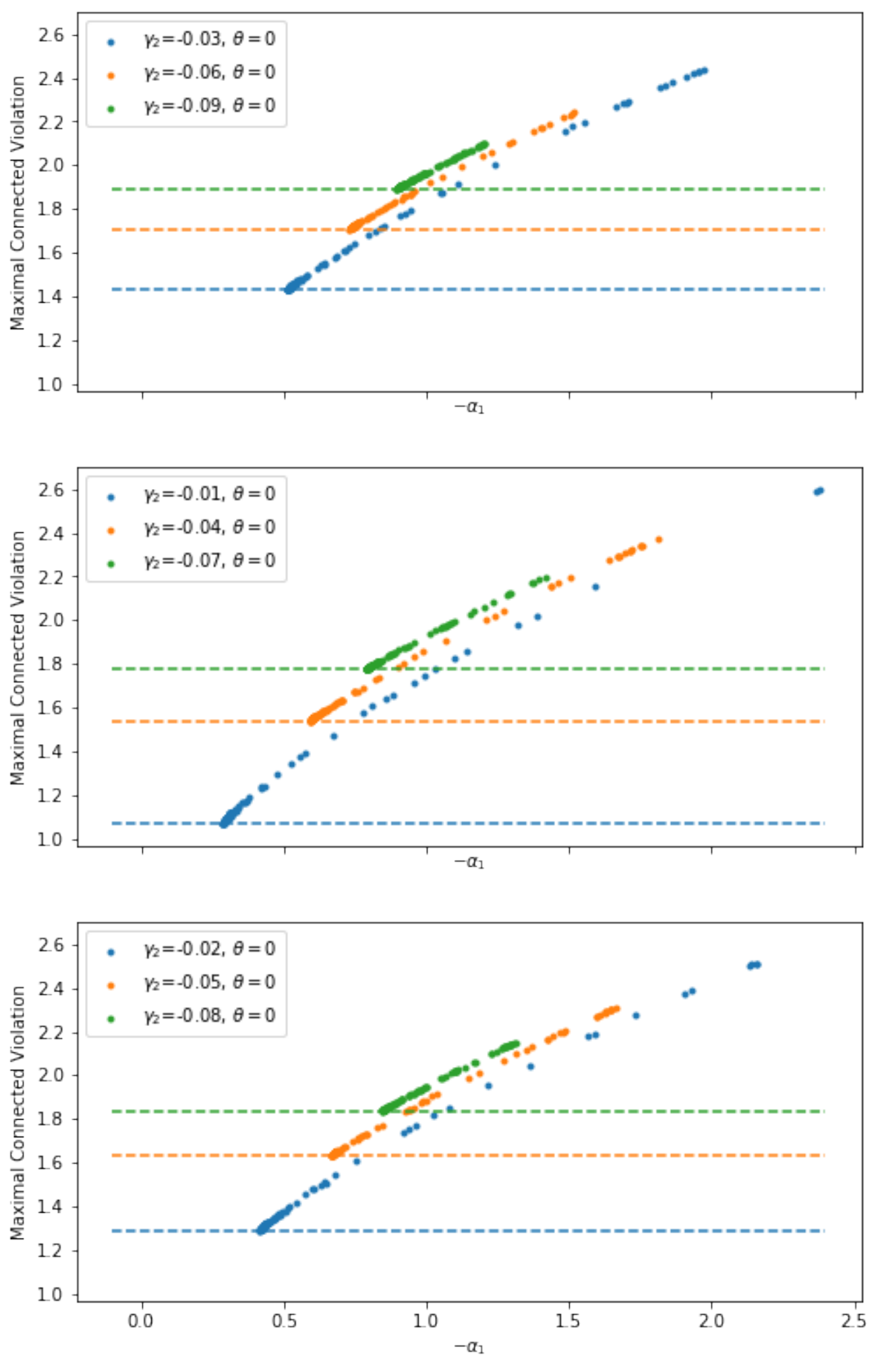}
\includegraphics[width=0.32\textwidth]{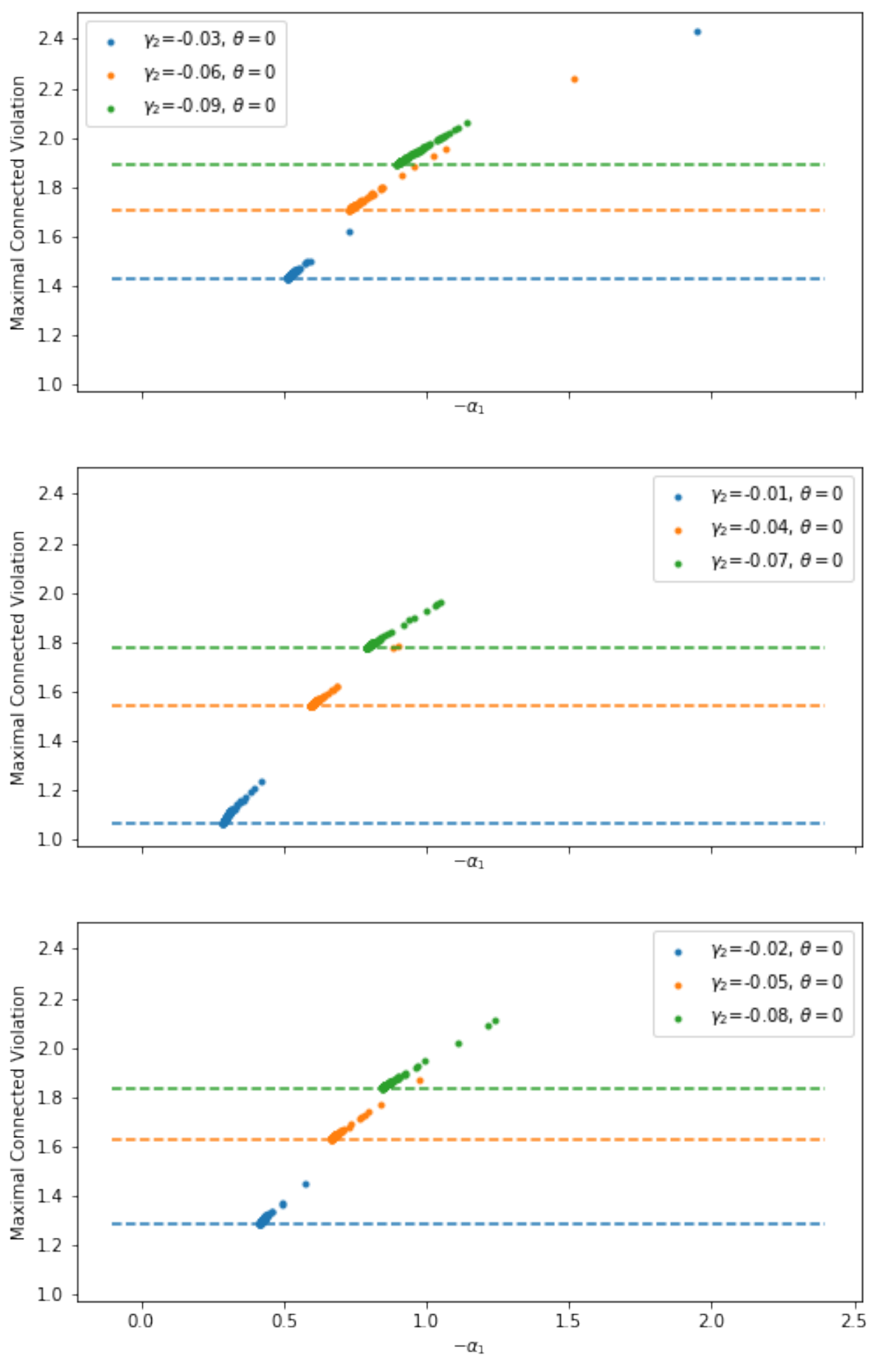} 
\includegraphics[width=0.32\textwidth]{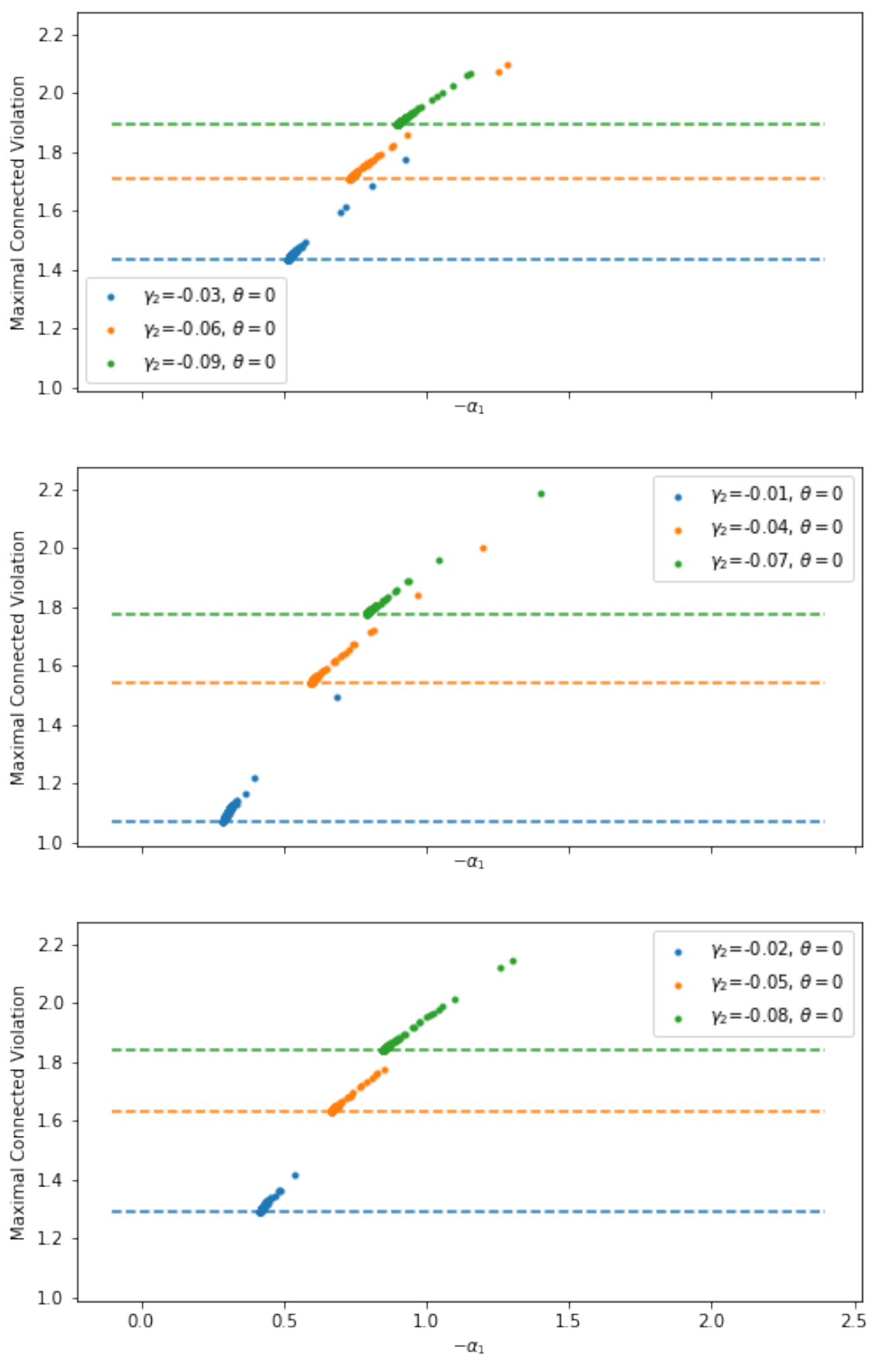}
\caption{We show the maximal connected violation $\gamma_c$ versus $-\alpha_1$. 
The dot lines correspond to the separable state. 
We fix $\gamma_2$ and $\theta$. 
In each classified class, the separable state is represented by the lowest value of $\gamma_c$. 
The result also shows the monotonic increasing behavior. 
The left, middle, and right plots correspond to the reduced density matrices, $\rho_{12}$, $\rho_{13}$, and $\rho_{23}$, respectively. 
} 
\label{cvio-a1}
\end{center}
\end{figure}

\begin{figure}[!htb]
\begin{center}
\includegraphics[width=0.32\textwidth]{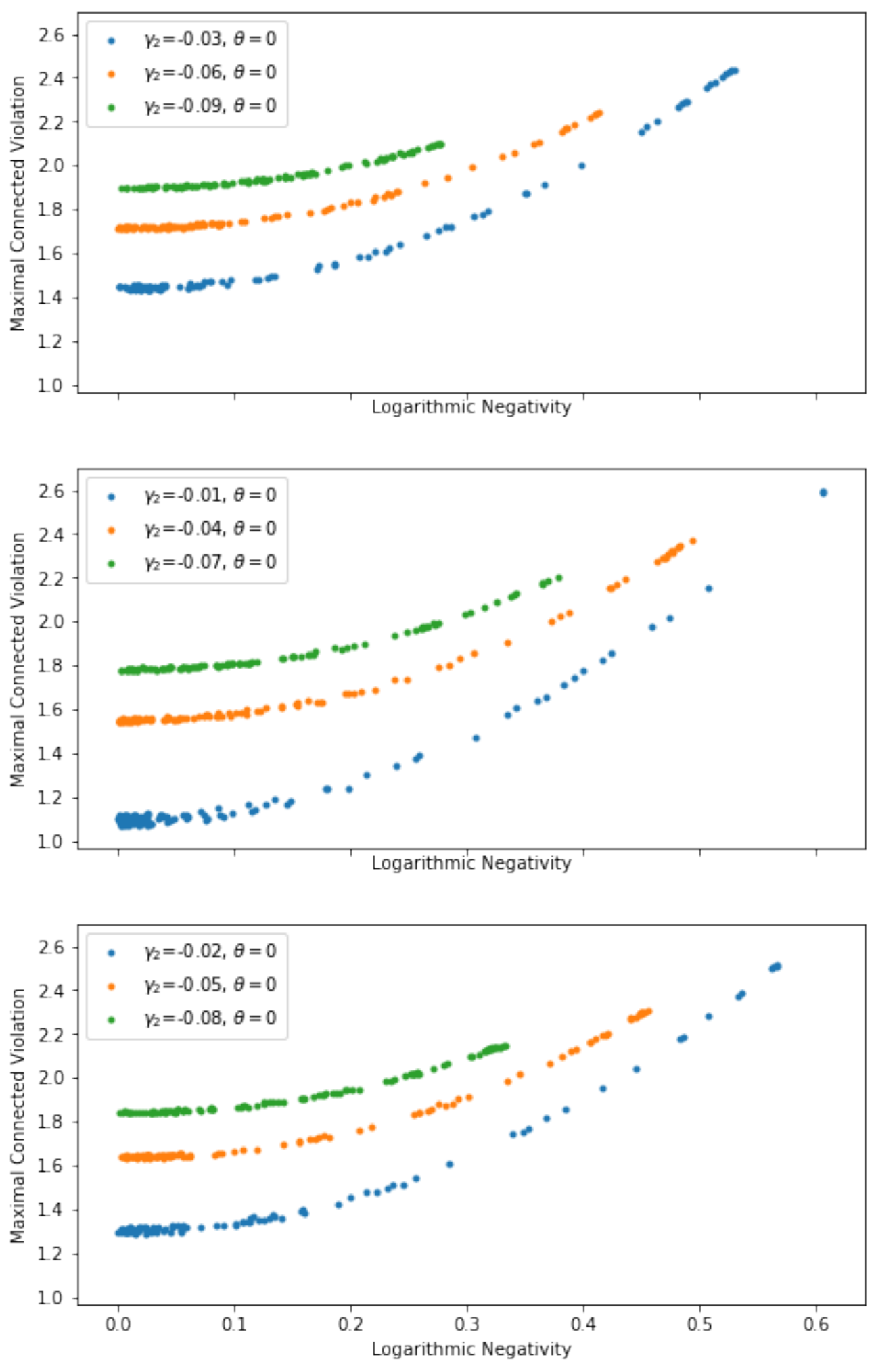}
\includegraphics[width=0.32\textwidth]{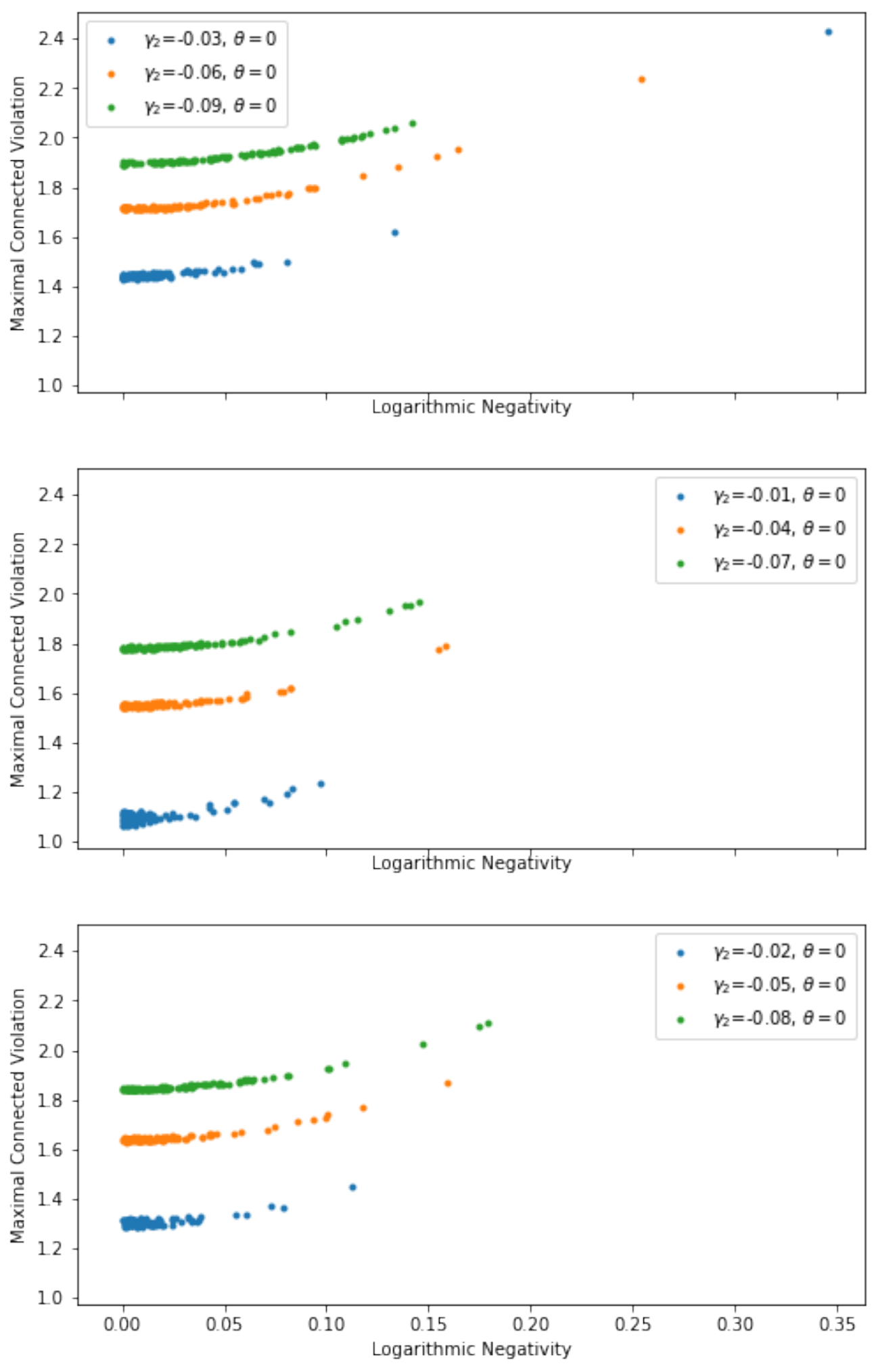} 
\includegraphics[width=0.32\textwidth]{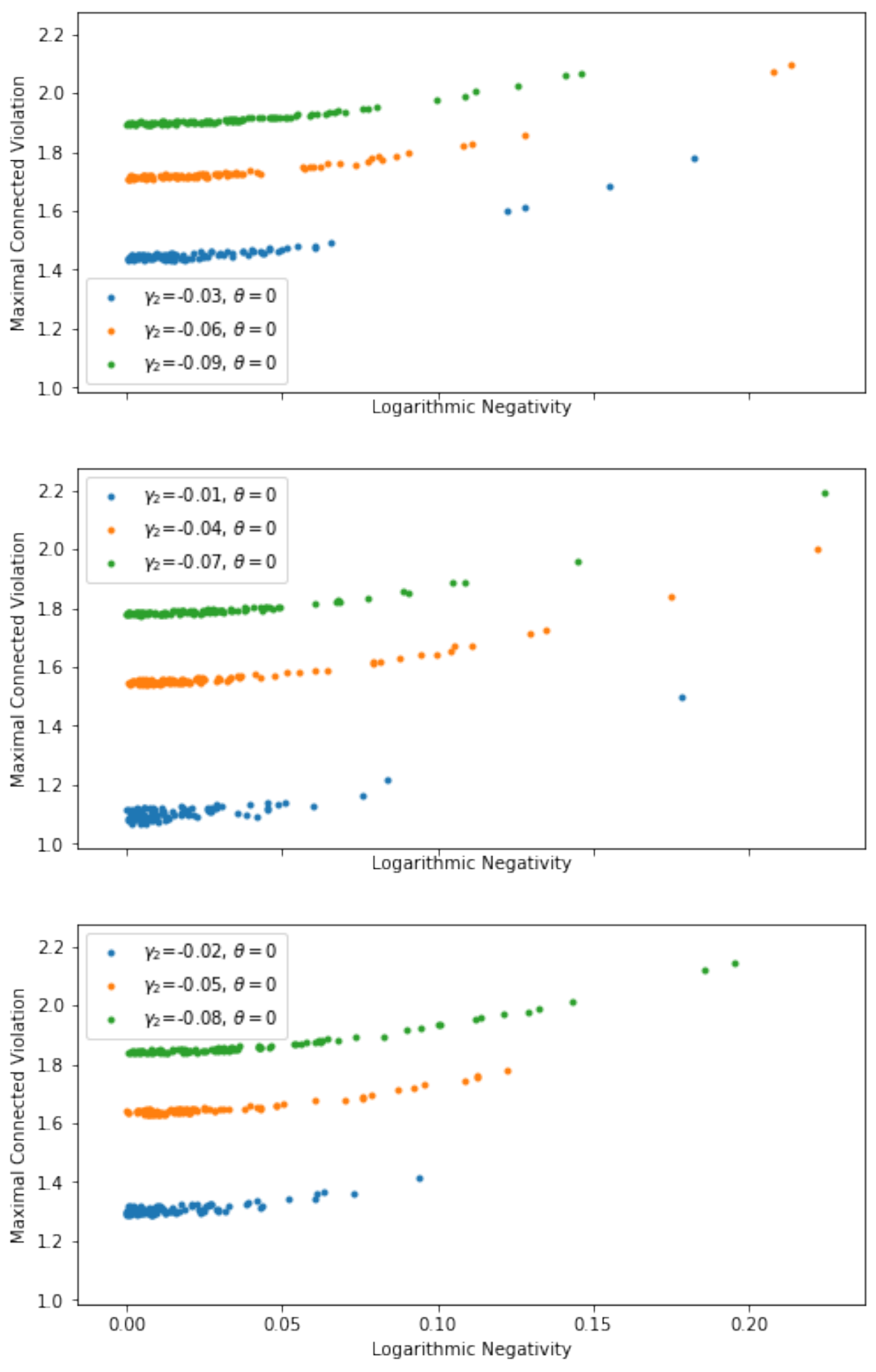}
\caption{We show the maximal connected violation $\gamma_c$ versus the logarithmic negativity. 
We then fix $\gamma_2$ and $\theta$. 
The logarithmic negativity does not have a one-to-one correspondence to $\gamma_c$.  
Therefore, the logarithmic negativity loses the monotonic increasing behavior. 
The left, middle, and right plots correspond to the reduced density matrices, $\rho_{12}$, $\rho_{13}$, and $\rho_{23}$, respectively. 
} 
\label{cvio-neg-a1}
\end{center}
\end{figure}

\begin{figure}[!htb]
\begin{center}
\includegraphics[width=0.32\textwidth]{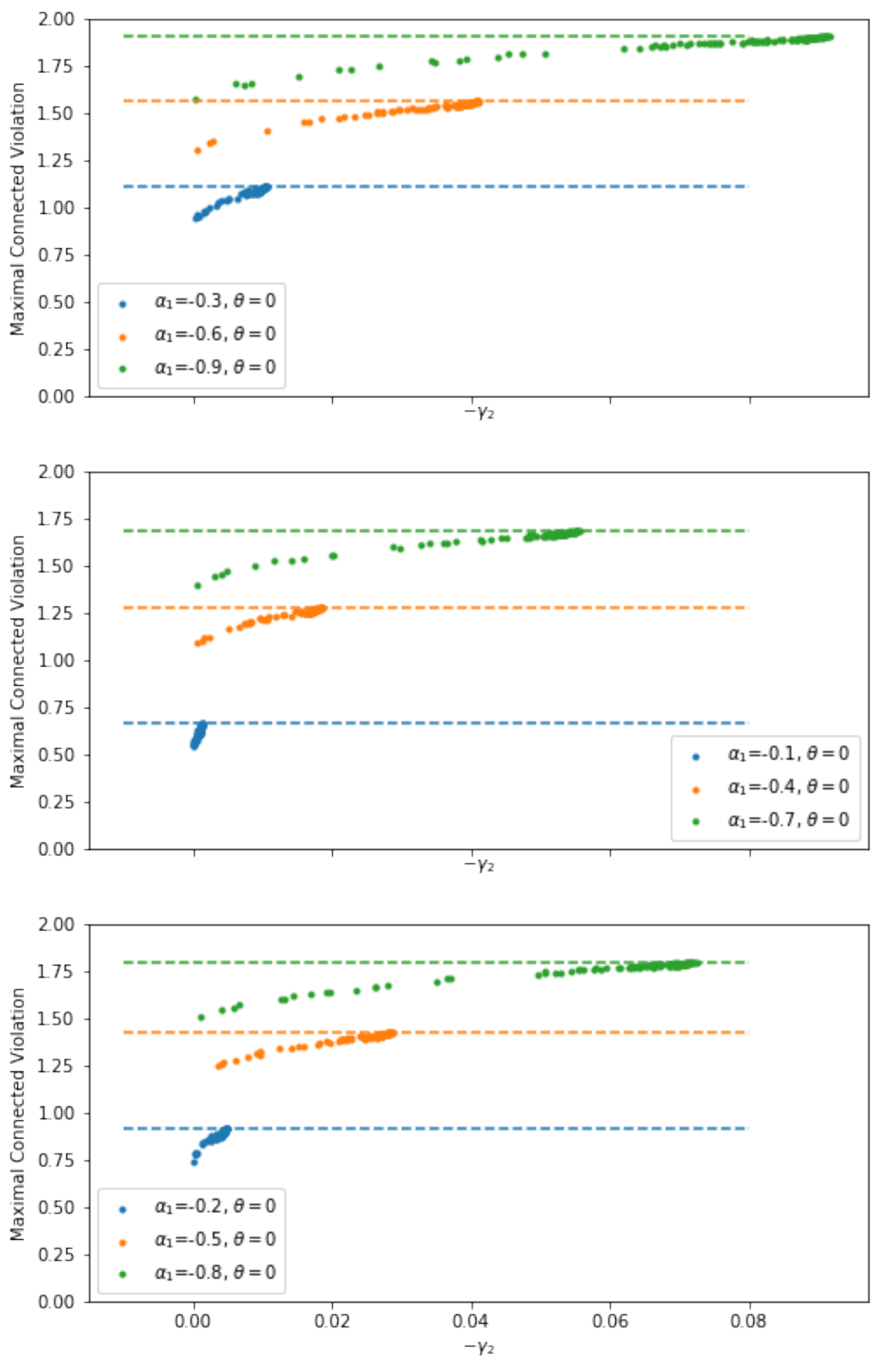}
\includegraphics[width=0.32\textwidth]{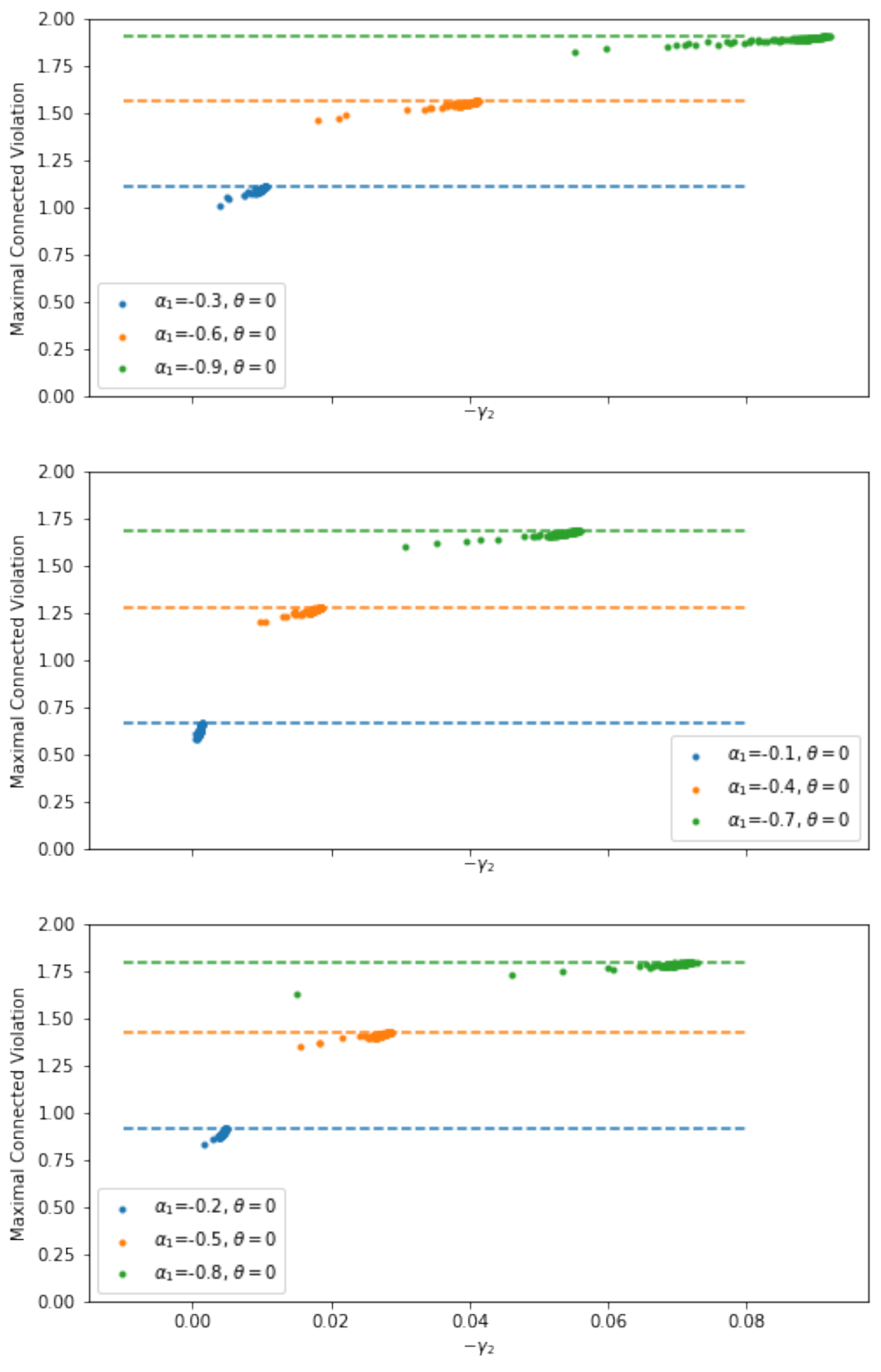} 
\includegraphics[width=0.32\textwidth]{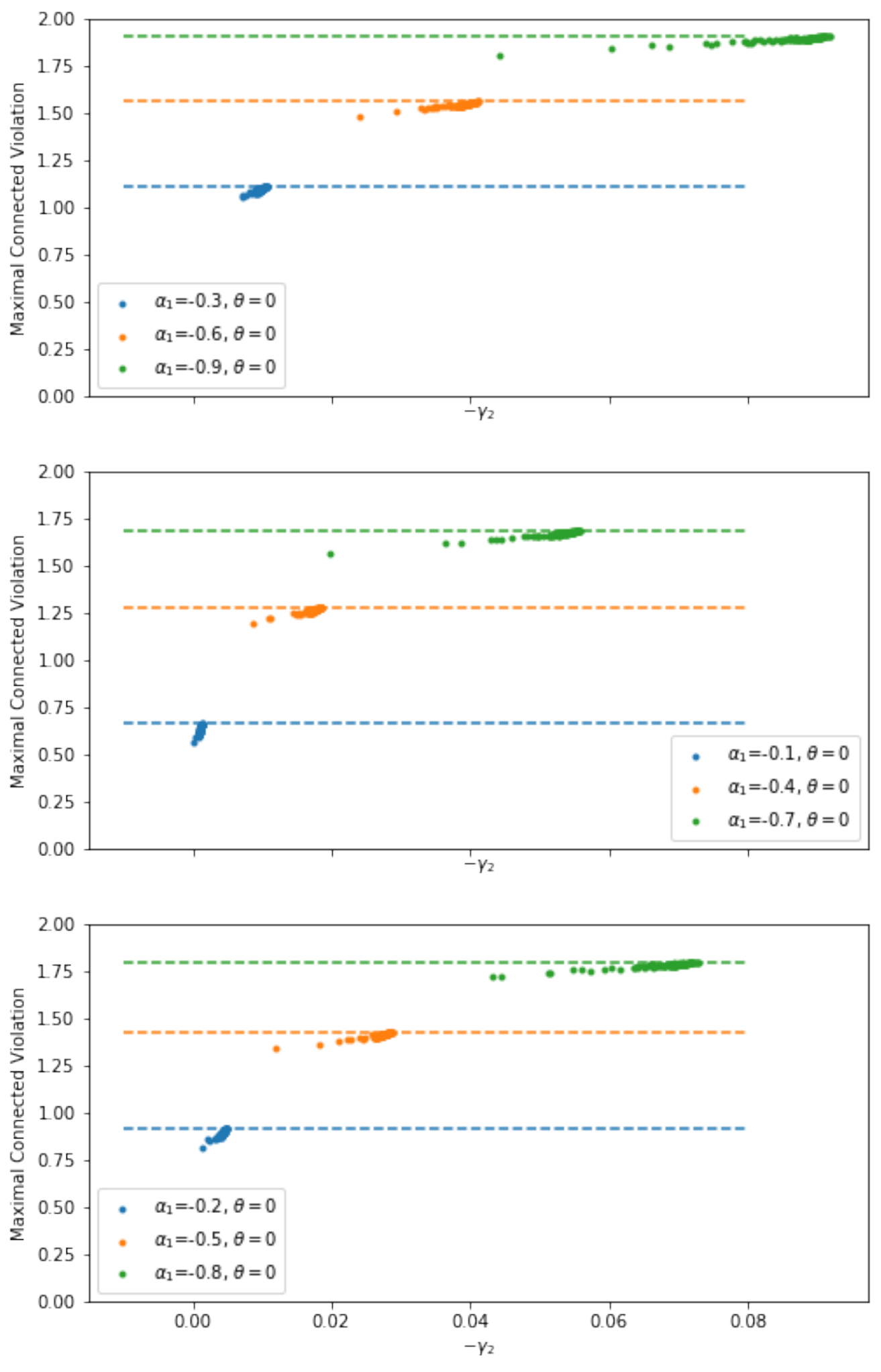}
\caption{We show the maximal connected violation $\gamma_c$ versus $-\gamma_2$. 
The dot lines correspond to the separable state. 
We fix $\alpha_1$ and $\theta$. 
In each classified class, the highest value of $\gamma_c$ corresponds to the separable state. 
Therefore, the choice of fixing is not suitable for diagnosing Quantum Entanglement. 
The left, middle, and right plots correspond to the reduced density matrices, $\rho_{12}$, $\rho_{13}$, and $\rho_{23}$, respectively. 
} 
\label{cvio-g2}
\end{center}
\end{figure}

\noindent 
The $E_1=0$  implies the separable state for the $\rho_{12}$ case. 
The maximal connected violation is $\gamma_c=2\sqrt{-\alpha_1}$ corresponding to: 
\bea
\alpha_1=-(E_2^2+E_4^2)(E_3^2+E_4^2), \ \alpha_2=\alpha_3=0.
\eea 
The separable state, in general, has a connected correlation \cite{Werner:1989zz}. 
One cannot use $\gamma_c=0$ to determine the separability. 
We propose the classification from two parameters, $\gamma_2$ and $\theta$ (based on the monotonic behaviors). 
The separable state corresponds to $\alpha_2=\theta=0$. 
When fixing $\gamma_2$, one needs to increase $\alpha_1$ for the non-zero $\alpha_2$. 
Therefore, the separable state always has the lowest value of $\gamma_c$ in each classified case. 
In other words, we can use $\gamma_c$ to measure the amount of Quantum Entanglement when fixing $\gamma_2$ and $\theta$.  
We demonstrate the result in Fig. \ref{cvio-a1}. 
The standard mixed state measure, logarithmic negativity $E_N\equiv\ln (2N +1)$ \cite{Peres:1996dw}, shows increasing behavior but is not monotonic in Fig. \ref{cvio-neg-a1}.
Hence our proposal is different from $E_N$ and successfully quantifies Quantum Entanglement. 
If we fix $\alpha_1$ and $\theta$, the same reason implies that the separable state lies in the highest $\gamma_c$ shown in Fig. \ref{cvio-g2}. 
Our study respects the three-qubit pure state \cite{Guo:2021ghs}. 
The choice for fixing parameters has ambiguity in the three-qubit state \cite{Guo:2021ghs}. 
The mixed separable state restricts the choice of fixing parameters.

\section{Locality and Quantum Entanglement} 
\label{sec:5}
\noindent 
Because the experimental result violates Bell's theorem \cite{Bell:1964kc}, people expect the generation of non-locality \cite{Clauser:1969ny} from Quantum Entanglement. 
A simple way to avoid the issue is to break the locality condition of Bell's theorem. 
In this context, we are exploring an alternative method to ensure that the locality is given due consideration. 
We employ the singular value decomposition to diagonalize $R$ (or change the basis). 
The quantum correlator becomes: 
\bea
c_{\mathrm{Q}}(\vec{a}, \vec{b})&=&q_x\tilde{a}_x\tilde{b}_x
+q_y\tilde{a}_y\tilde{b}_y
+q_z\tilde{a}_z\tilde{b}_z
\nn\\
&=&\int_0^1 d\lambda\ \vec{F}(\tilde{a}, \lambda)\cdot\vec{F}(\tilde{b}, \lambda)
, 
\eea
where $\tilde{a}_x$ is the $x$-component of unit vector $\vec{\tilde{a}}$ (on the new basis), 
\bea
\vec{F}(a, \lambda)=(\lambda^{k_x}a_x, \lambda^{k_y}a_y, \lambda^{k_z}a_z); \ k_j=\frac{1-q_j}{2q_j}. 
\eea   
The $\lambda$ is the hidden variable with a uniform distribution, $\int dP(\lambda)\sim\int_0^1d\lambda$. 
Therefore, we do not violate the freedom of choice. 
Another measurable observable is spin $S_a$ for the operator $\vec{a}\cdot\vec{\sigma}$. 
Therefore, we should determine the quantum correlator by summing all spins 
\bea
c_{\mathrm{Q}}(\vec{a}, \vec{b})=\sum_{S_a, S_b=\pm 1}S_aS_b P(S_a, S_b| \vec{a}, \vec{b}), 
\eea
where $P(S_a, S_b| \vec{a}, \vec{b})$ is the conditional probability distribution of spins. 
Hence we can rewrite $P(S_a, S_b| \vec{a}, \vec{b})$ as 
\bea
P(S_a, S_b| \vec{a}, \vec{b})=\int_0^1d\lambda\ P(S_a, S_b| \vec{a}, \vec{b}, \lambda). 
\eea
We modify the locality condition of Bell's theorem to that 
\bea
P(S_a, S_b| \vec{a}, \vec{b}, \lambda)&=&P(S_a| \vec{a}, \lambda)P(S_b| \vec{b}, \lambda)
\nn\\
\Longrightarrow 
P(S_a, S_b| \vec{a}, \vec{b}, \lambda)&=&\sum_{k=1}^3P_k(S_a| \vec{a}, \lambda)P_k(S_b| \vec{b}, \lambda). 
\nn\\
\eea
The probability is also normalizable by summing all spins and components 
\bea
\sum_{S_a, S_b=\pm 1}\sum_{k=1}^3 P_k(S_a| \vec{a}, \lambda)P_k(S_b, \vec{b}, \lambda)=1.  
\eea
One choice of probability distribution satisfying all conditions is: 
\bea
P_j(S_a=1| \vec{a}, \lambda)&=&\frac{\sqrt{3}+3\lambda^{k_j} a_j}{6}; 
\nn\\ 
P_j(S_a=-1| \vec{a}, \lambda)&=&\frac{\sqrt{3}-3\lambda^{k_j}a_j}{6}. 
\eea
We can violate Bell's theorem because the vector observable $\vec{F}$ is not in the setup of Bell's theorem. 
The vector observable implies the necessity of a simultaneous observation (for all components of $\vec{F}$). 
The $c_{\mathrm{Q}}(\vec{a}, \vec{b})$ still respects the factorization. 
Therefore, this local hidden variable model should show that the locality condition of Bell's theorem is too restricted.  
We can also formulate the discrete variable, components of a probability distribution or $\vec{F}$, as $\lambda$ but loses the freedom of choice: 
\bea
&&
P(S_a, S_b| \vec{a}, \vec{b}, \lambda)
\nn\\
&=&
\sum_{k=1}^3P(S_a, S_b|\vec{a}, \vec{b}, \lambda, k)P(k|\vec{a}, \vec{b}, \lambda)
\nn\\
&=&
\sum_{k=1}^3 P(S_a| \vec{a}, \lambda, k)P(S_b| \vec{b}, \lambda, k)P(k| \vec{a}, \vec{b}, \lambda), 
\eea
where 
\bea
P(S_a| \vec{a}, \lambda, k)&\equiv&\frac{P_k(S_a| \vec{a}, \lambda)}{\sqrt{P(k| \vec{a}, \vec{b}, \lambda)}}, 
\nn\\
P(k| \vec{a}, \vec{b}, \lambda)&\equiv& \sum_{S_a, S_b=\pm 1} P_k(S_a| \vec{a}, \lambda)P_k(S_b| \vec{b}, \lambda).   
\eea
The new probability distribution does not break the normalizability 
\bea
\sum_{S_a, S_b=\pm 1}P(S_a| \vec{a}, \lambda, k)P(S_b| \vec{b}, \lambda, k)=1. 
\eea 
Hence we realize the local hidden variable model and show its existence. 
\\

\noindent
Now $k$ becomes a new local hidden variable. 
Therefore, we apply the locality condition of Bell's theorem to the probability distribution 
\bea
P(S_a, S_b|\vec{a}, \vec{b}, \lambda, k)=P(S_a| \vec{a}, \lambda, k)P(S_b| \vec{b}, \lambda, k).
\eea 
The $k$ does not satisfy the freedom to choose, $P(k|\vec{a}, \vec{b}, \lambda)\neq P(k)$. 
Hence preserving the locality needs to pay for the freedom to choose. 
\\

\noindent
The quantification of Quantum Entanglement is through a connected correlator. 
We can apply this local hidden variable model to $c_{\mathrm{C}}(\vec{a}, \vec{b})$. 
Therefore, our demonstration proves that non-locality is not a necessary consequence of Quantum Entanglement. 

\section{Outlook}
\label{sec:6}
\noindent 
We research Quantum Entanglement and its relationship with connected correlation and local hidden variable theory. 
The use of a connected correlation matrix to quantify Quantum Entanglement is a notable approach. 
This matrix enables the measurement and understanding of correlations between entangled particles. 
It is possible for separable states, where each particle's state can be described independently, to have complex correlations between them. 
This creates a challenge in distinguishing between entangled states and separable states that have significant connected correlations. 
To tackle this issue, we have proposed a classification system that can differentiate between entangled states and separable states with non-trivial connected correlations. 
This scheme aids in identifying and quantifying the degree of entanglement in systems exhibiting such correlations. 
Quantifying the degree of entanglement in systems with significant connected correlations is indeed a critical step in understanding the behavior of Quantum Entanglement.   
\\

\noindent
Tracing out one qubit from a three-qubit system is a common technique in quantum information theory. 
This process allows us to obtain a reduced density matrix for the remaining qubits, which is useful for studying the entanglement properties of the original two-qubit system. 
The connected correlator is a measure of quantum correlation. 
We suggest that using the connected correlator on a two-qubit mixed state helps quantify the amount of entanglement present. 
The fact that the quantum correlator does not work indicates that quantum entanglement is the source of the connected correlation. 
While a general density matrix for a two-qubit system has 16 elements, not all of these elements are independent. 
Tracing out one qubit from a three-qubit pure state enables the characterization of two-qubit mixed states. 
This process is essential for understanding the dynamics of entanglement in multi-qubit systems. 
Studying single-qubit unitary transformations is crucial for understanding how operations on individual qubits affect the overall entanglement structure of multi-qubit systems. 
Purification involves embedding a mixed state into a larger pure state. 
Using a four-qubit pure state to obtain all possible two-qubit mixed states through purification highlights the versatility of this technique in representing quantum states. 
Understanding entanglement measures resulting from single-qubit unitary transformations is important for comprehending the properties and behaviors of multi-qubit systems, especially those involving four-qubit systems.
\\ 

\noindent 
Determining whether a given state is entangled or separable is a fundamental question in quantum information theory. 
While it remains computationally challenging in general, certain classes of states can be solved efficiently. 
Ongoing research focuses on finding efficient and general methods to solve this problem for arbitrary quantum states. 
Various approaches and techniques, such as semidefinite programming, entanglement witnesses, and numerical optimization methods, are being explored to tackle this problem and improve our understanding of entanglement in quantum systems.
We observed that the connected correlation is only relevant to the two largest eigenvalues of $R_c^TR_c$ for any $n$-qubit states. 
Our case showed that a separable state at least has two zero eigenvalues. 
The statement helps distinguish the separable and entangled states. 
Then we should provide one efficient way to solve "the separability problem". 
\\

\noindent
Our study explored the implications of local hidden variable theory for the connected correlator and its relation to Bell's theorem \cite{Bell:1964kc} and experimental results \cite{Clauser:1969ny}. 
By considering a factorization of the observable into two vectors instead of two scalars, we have not found violations of the locality assumption in Bell's theorem. 
This approach is taken to explore violations of the locality assumption in Bell's theorem. 
The results indicate that with the vector factorization approach, violations of the locality assumption in Bell's theorem are not observed. 
The findings suggest that Quantum Entanglement, in this context, does not lead to non-locality. 
Experimental tests can be conducted using low-dimensional models that can be realized in the laboratory. 
These tests aim to determine the consistency between the vector observable and experimental outcomes. 
The use of the vector observable approach in experiments is suggested to provide insights into the relationship between theoretical models and empirical observations.

\section*{Acknowledgments}
\noindent 
We thank Ki-Seok Kim, Yuya Kusuki, and Masaki Tezuka for their helpful discussion. 
CTM would thank Nan-Peng Ma for his encouragement. 
XG acknowledges the Guangdong Major Project of Basic and Applied Basic Research No. 2020B0301030008 and NSFC Grant No.11905066.
CTM acknowledges the DOE grant (Grant No. GR-024204-0001); 
YST Program of the APCTP; 
Post-Doctoral International Exchange Program (Grant No. YJ20180087); 
China Postdoctoral Science Foundation, Postdoctoral General Funding: Second Class (Grant No. 2019M652926); 
Foreign Young Talents Program (Grant No. QN20200230017).


\end{document}